%% file: eprint.tex
\documentclass[12pt]{article}
\usepackage{graphicx}
\usepackage{subcaption}
\usepackage{xspace}
\usepackage{cite}
\input command.tex

\def\lbnl{Nuclear Science Division, Lawrence Berkeley National Laboratory, Berkeley CA 94720 USA}

\def\Title#1{\begin{center} {\Large #1 } \end{center}}
\def\Author#1{\begin{center}{ \sc #1} \end{center}}
\def\Address#1{\begin{center}{ \it #1} \end{center}}

\newenvironment{Abstract}{\begin{quotation}  }{\end{quotation}}
\newenvironment{Presented}{\begin{quotation} \begin{center} 
             PRESENTED AT\end{center}\bigskip 
      \begin{center}\begin{large}}{\end{large}\end{center} \end{quotation}}
\def\Acknowledgements{\bigskip  \bigskip \begin{center} \begin{large}
             \bf ACKNOWLEDGEMENTS \end{large}\end{center}}

\input econfmacros.tex

\begin{document}
\begin{titlepage}

\vfill
\Title{Dipion and dikaon photoproduction in ultra-peripheral Pb-Pb collisions with ALICE}
\vfill
\Author{Abdennacer Hamdi on behalf of the ALICE Collaboration}
\Address{\lbnl}
\vfill
\begin{Abstract}
      High energy photons originating from the electromagnetic field of ultrarelativistic lead nuclei can interact with the other lead ion. These reactions are studied in the ultra-peripheral heavy ion collisions to probe the physics of strong interactions. The analysis of dipion and dikaon photoproduction was carried out using the ALICE 2015 \PbPb data at center-of-mass energy \five, and 2017 \XeXe data at $\sqrt{s}~=~5.44$~TeV. We present the $\rho^0$ meson and direct \pip\pim measurements, as well as the prospects of studying exclusive \kk photoproduction, with sensitivity towards higher dikaon invariant mass above the $\phi(1020)$ threshold.
\end{Abstract}
\vfill
\begin{Presented}
      DIS2023: XXX International Workshop on Deep-Inelastic Scattering and
Related Subjects, \\
Michigan State University, USA, 27-31 March 2023 \\
     \includegraphics[width=9cm]{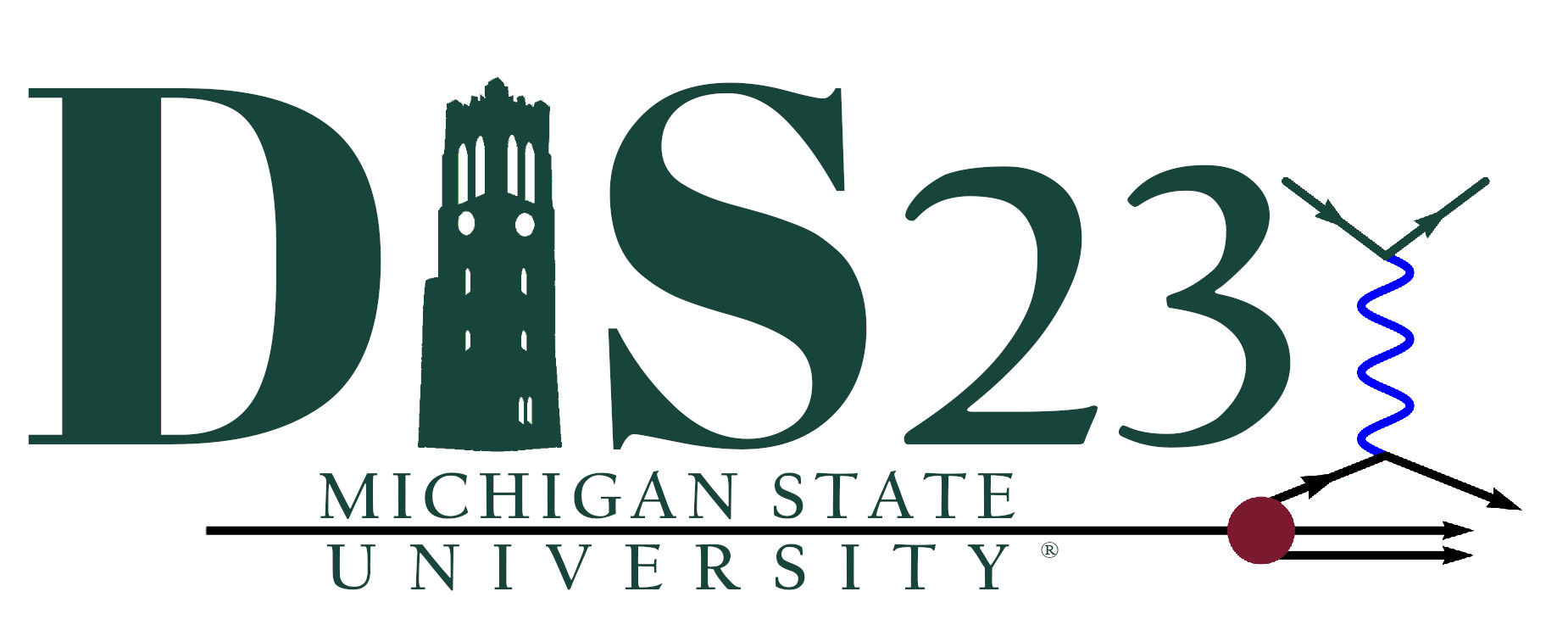}
\end{Presented}
\vfill
\end{titlepage}
\def\thefootnote{\fnsymbol{footnote}}
\setcounter{footnote}{0}

\section{Introduction}

Photon-induced interactions can be produced abundantly in Ultra-Peripheral Collisions (UPC) of relativistic heavy ions~\cite{Bertulani, Klein1, Contreras}, with the very intense photon beams reaching the energy frontiers at the LHC. At UPC the impact parameter is greater then the sum of the two colliding nuclei radii, and due to the short range nature of Quantum Chromo-Dynamics (QCD), the hadronic interactions are suppressed.

The photons can fluctuate to a quark-antiquark dipole~\cite{Bauer} and scatters off the entire nucleus target coherently, or on some nucleon sites incoherently. The vector meson is then produced with the same quantum number, $J^{PC} = 1^{--}$, as the photon, which then decays to a final state dihadrons or dileptons, e.g.: dipions and dikaons. In addition to the vector meson production, the photons can fluctuate directly to the dihadrons, leaving two undistinguishable final states, which leads to an interference of the direct and resonant production modes.

The photoproduction of vector mesons is expected to be sensitive to the gluon distribution in nuclei~\cite{Ryskin,Eskola}. Allowing to study nuclear shadowing effects~\cite{Armesto}, suppression of the nuclear form factor of nuclei with respect to the superposition of its nucleons, both at low Bjorken-$x$ and low transferred 4-momentum $Q^2$.

The strong electromagnetic fields accompanying the colliding ions could give rise to further photon exchanges between the nuclei. This leads to nuclear fragmentation of one or both the excited ion targets, which is accompanied by neutron emission at beam rapidities. In UPC, the photons could be emitted from either nucleus, leading to a two-fold ambiguity on the photon energy~\cite{Klein2}. The measurement of vector mesons cross section in different neutron emission classes at beam rapidities has been proposed as a tool to disentangle the two photon energies~\cite{Baltz,Guzey}.

\section{Experimental Setup and Event Selection}

The ALICE detector, designed with wide kinematic coverage, is optimized to study the high
particle-multiplicity environment of ultra-relativistic heavy-ion collisions, full detector description can be found in~\cite{Abelev,Aamodt}. For the photoproduction of pion and kaon pairs in UPC events, few sub-detectors were used. Mainly, the Inner Tracking System (\ITS), where the pair tracks traversing its silicon layers in a near back-to-back topology where selected. The pions/kaons are identified with their energy loss (\dEdx) in the Time Projection Chamber (\TPC), more details on kaon identification is discussed in section~\ref{sec:dikaon}. Both tracking systems are placed in a large solenoid magnet of 0.5 T, and offer a pseudorapidity coverage \etarange 0.9 and full azimuth.

In addition, a dedicated trigger is used to select UPC events, involving vetoes on any activity in forward \VZEROA ($2.8<\eta<5.1$) and ADA ($4.7<\eta<6.3$) and backward \VZEROC ($-3.7<\eta<-1.7$) and ADC ($-6.9<\eta<-4.9$) scintillators.

Two Zero Degree Calorimeters (\ZDC) are used to detect neutrons from the electromagnetic dissociation of the target nuclei. They are located at $\pm 112.5$ m from the interaction point and have a coverage of $|\eta|>8.8$.

\section{Coherent ${\bf \rho^0}$ Photoproduction}

\begin{figure}[htb]
      \centering
      \includegraphics[height=2.5in]{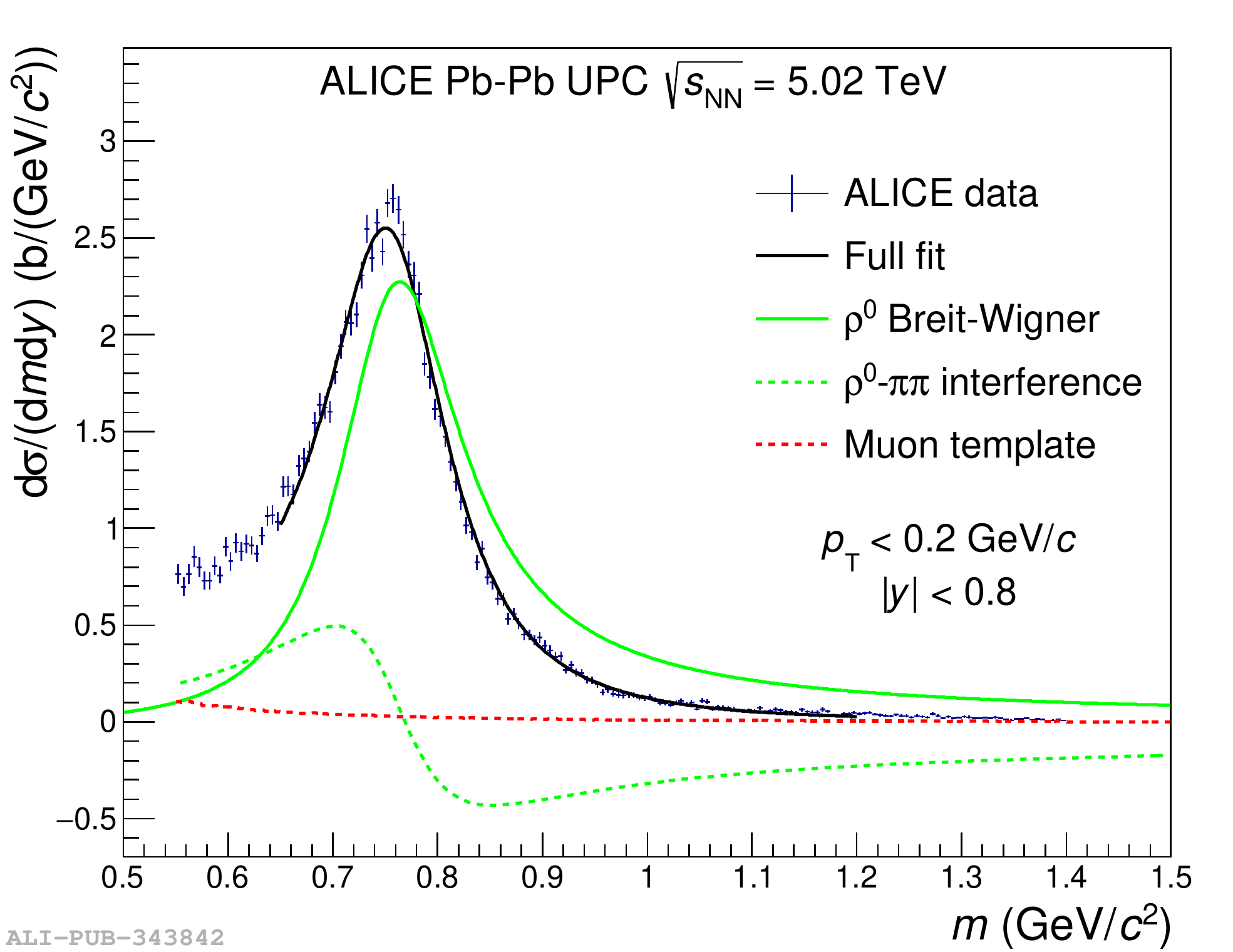}
      \caption{Mass-dependent cross section of exclusive dipions in the coherent pair range \pt $<$ 0.2 \GeVc, well described by a Breit--Wigner fit for the $\rho^0$ and direct \pip\pim, the fit includes background from $\gamma \gamma \rightarrow \mu^+\mu^-$, figure taken from~\cite{rho_pb}.}
      \label{fig:dipion_invariant_mass}
\end{figure}

The measurement of mass-dependent dipion photoproduction cross section, after event selection, shows a clean $\rho^0$ peak in the pair invariant mass distribution in Fig.~\ref{fig:dipion_invariant_mass}. The coherent \pip\pim pairs were selected in the transverse momentum range \pt $<$ 0.2 \GeVc and rapidity $|y|<$ 0.8. The $\rho^0$ resonance and direct \pip\pim are well described by the mass-dependent width Breit--Wigner fit. A template for the remaining small background contribution, $\gamma \gamma \rightarrow \mu^+\mu^-$, is added to the fit. Similar to \PbPb UPC, the exclusive $\rho^0$ cross section was also measured for the first time in $\gamma \mathrm{Xe} \rightarrow \rho^0 \mathrm{Xe}$ process, with similar $\rho^0$ resonance shape observation in the dipion invariant mass. The contribution of the direct \pip\pim to the $\rho^0$ resonance production cross section ratio is found to be consistent for both $\gamma$Pb and $\gamma$Xe with $0.57 \pm 0.01$ (stat.) $\pm 0.02$ (syst.) (\GeVmass)$^{-1/2}$ in Pb and $0.58 \pm 0.04$ (stat.) $\pm 0.03$ (syst.) (\GeVmass)$^{-1/2}$ in Xe.

\begin{figure}[htb]
      \centering
      \includegraphics[height=2.3in]{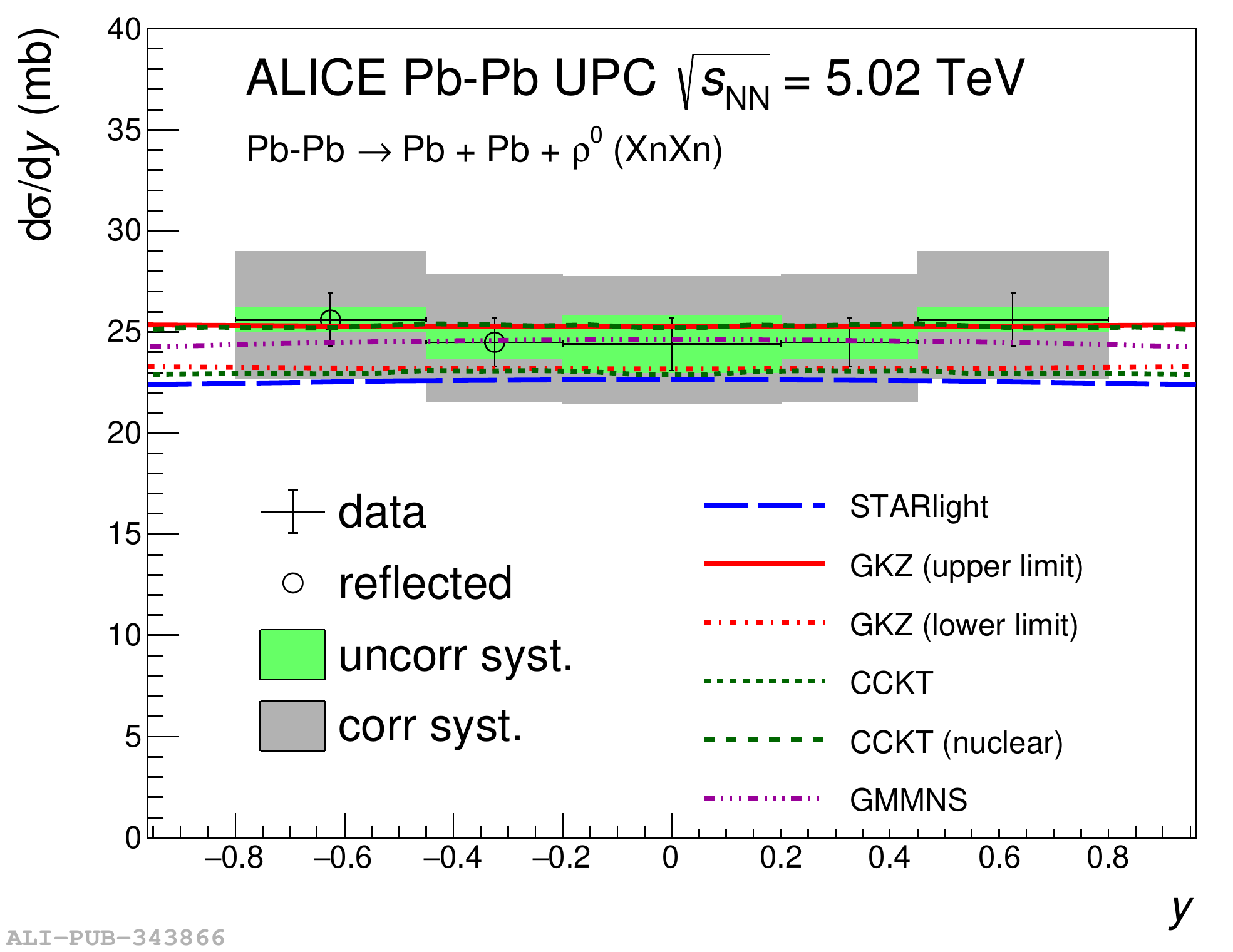}
      \includegraphics[height=2.3in]{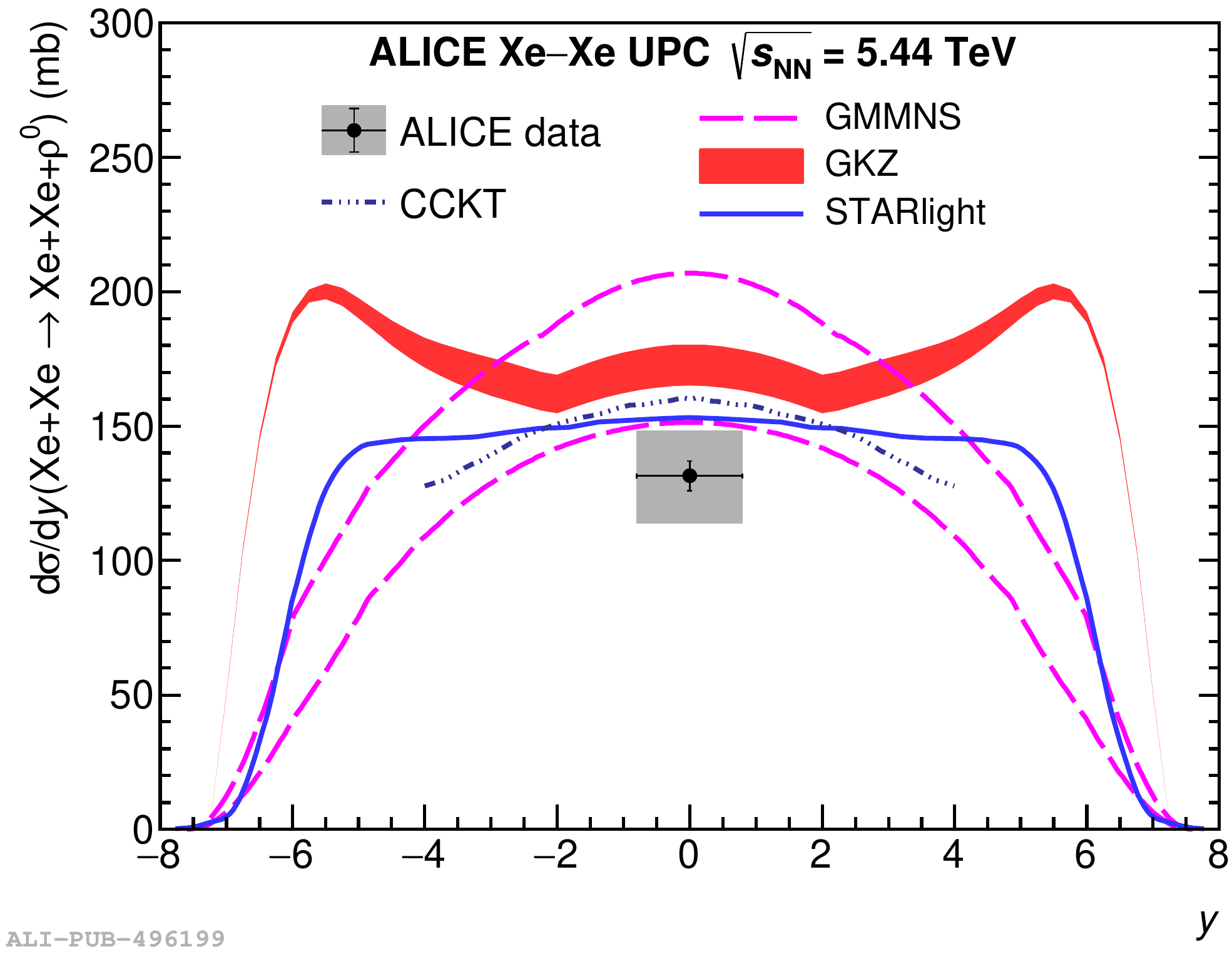}
      \caption{Rapidity-dependent cross section for dipion photoproduction in UPC for (Left) \PbPb and (right) \XeXe, figures taken from~\cite{rho_pb, rho_xe}.}
      \label{fig:rho_dxsecdy}
\end{figure}

The rapidity-dependent $\rho^0$ photoproduction cross section in \PbPb and \XeXe at UPC is compared to different model predictions (fig.~\ref{fig:rho_dxsecdy}). STARlight is based on parameterized HERA data on exclusive photoproduction on proton, and a Glauber-like eikonal formalism for nuclear targets~\cite{starlight}. The GKZ prediction is based on the photon hadronic fluctuations interacting with the nuclei according to the Gribov-Glauber model of nuclear shadowing~\cite{gkz}. Variations of the GKZ prediction on the uncertainty of theory parameters are shown as upper and lower limit of the model. The CCKT model is based on the colour dipole model with the structure of the nucleon in the transverse plane described by high gluonic density (hot-spot)~\cite{cckt1, cckt2}. Two versions of the CCKT model are included, one with the hot-spot and one without (marked "nuclear"). The last model is GMMNS, which uses the colour dipole approach with amplitudes obtained from the IIM model~\cite{iim}, coupled to a Glauber prescription to go from the nucleon to the nuclear target~\cite{gmmns}.

The models describe fairly well the Pb data within uncertainties, in particular models based on nuclear shadowing (GKZ) and saturation (CCKT). For the Xe data, STARlight and CCKT as well as the lower band of GMMNS predictions are one standard deviation above data.

\section{Dikaon photoproduction in UPCs}
\label{sec:dikaon}

Similar to dipion photoproduction, dikaon final state may come from the $\phi(1020)\rightarrow K^+K^-$, or direct production. The two processes are undistinguishable and therefore interfere with each other. Since the kaons are much heavier than pions, we expect a much smaller dikaon photoporoduction cross section. Using current ALICE data, the dikaon production in UPC is sensitive to much higher $K^+K^-$ invariant mass, above the $\phi(1020)$ mass threshold. In this high mass region, very limited studies are conducted for dikaon photoproduction. 

After the initial two good tracks selection, and unlike the dipion sample, the dikaons are overpopulated with pions, with kaon:pion ratio $\sim$ 1 in 1000. This is resolved by using the particles energy loss \dEdx in \TPC, where the number of standard deviations of the measured \dEdx away from Bethe--Bloch expectation for each particle species is used, n$\sigma_i$ with $i = K, \pi, \mu, e$.

\begin{figure}[htb]
      \centering
      \begin{subfigure}[b]{0.5\textwidth}
          \includegraphics[width=\textwidth]{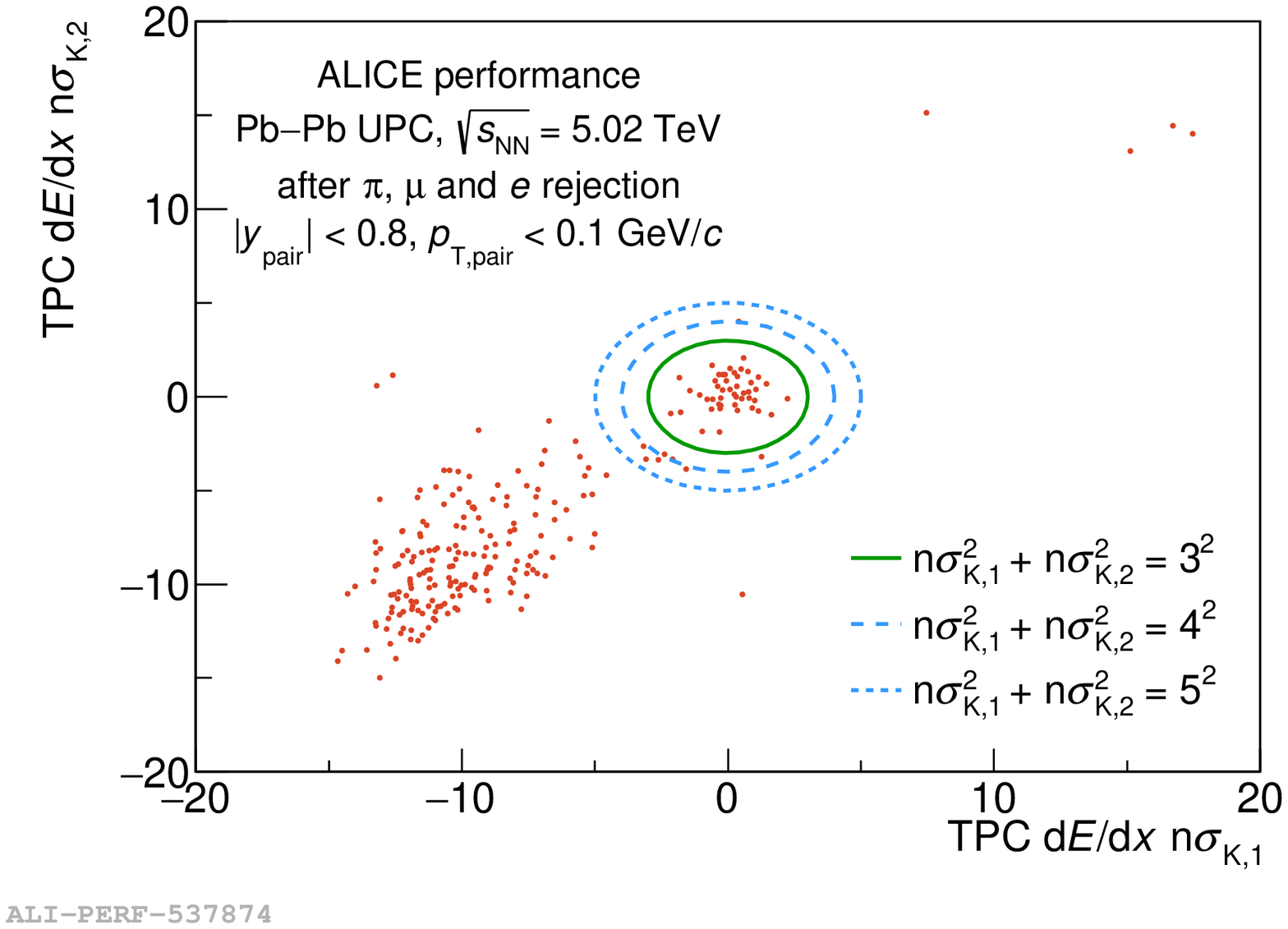}
          \caption{}
          \label{fig:dikaon.a}
      \end{subfigure}\hfill
      \begin{subfigure}[b]{0.5\textwidth}
          \includegraphics[width=\textwidth]{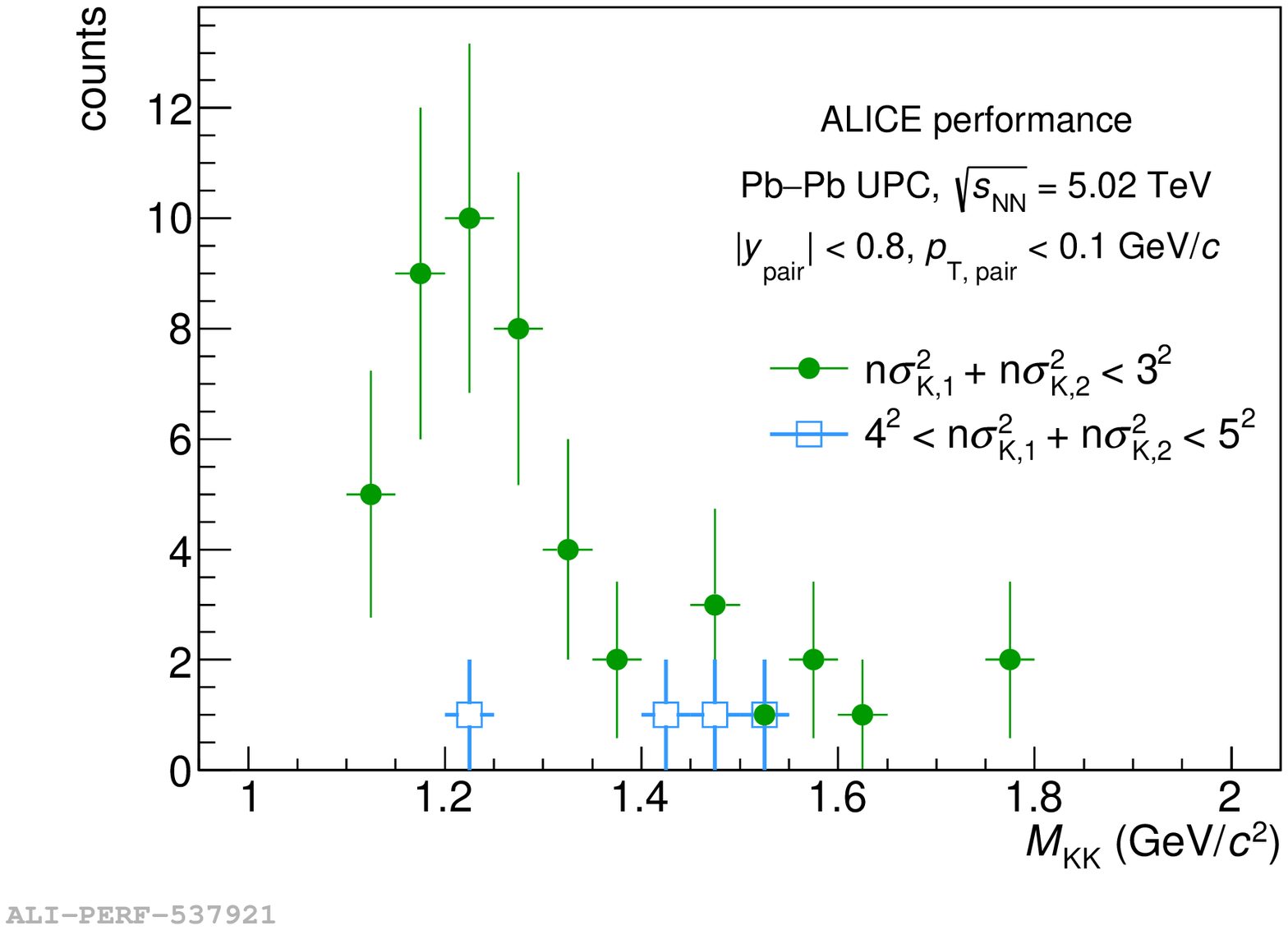}
          \caption{}
          \label{fig:dikaon.b}
      \end{subfigure}
      \caption{\label{fig:dikaon} (a) The TPC dE/dx in terms of the number of $\sigma$ away from the kaon hypothesis for all events after cuts to reject pions, muons and electrons, but without the cut $|n\sigma_K|$ $<$ 3. (b) Dikaon invariant mass for the signal (green) and background (blue) regions.}
\end{figure}

The contamination from pions and other lepton pairs are rejected by the \dEdx for each particle hypothesis, $|n\sigma_{(\pi, \mu, e)}|$ $>$ 2. Figure~\ref{fig:dikaon.a} shows the correlation between the two tracks in the pairs in terms of n$\sigma_K$ compatible with kaons, after $|n\sigma_{(\pi, \mu, e)}|$ $>$ 2. A clear kaon pairs signal is visible inside the green circle with a radius of 3$\sigma_K$ around the center point, while the remnant background is estimated outside the signal region with an equal area. The background region is estimated between two circles of 4$\sigma_K$ and 5$\sigma_K$ radius from the center, represented in the region between the two dashed blue lines. The latter is used to estimate the potential background inside the signal region. Figure~\ref{fig:dikaon.b} shows the dikaon invariant mass for the signal (green) and background (blue) regions. A clean dikaon signal with negligible background is observed in the \kk mass range [1.1, 1.4 \GeVmass], wile an increase in the background level is seen at higher mass. The coherent dikaon photoproduction cross section, with pair \pt $<$ 0.1 \GeVc and $|y| <$ 0.8, will be measured.

\Acknowledgements
This work is supported in part by the U.S. Department of Energy, Office of Science, Office of Nuclear Physics, under contract numbers DE-AC02-05CH11231.

\end{document}

%% file: command.tex
\newcommand{\XeXe}         {\mbox{Xe--Xe}\xspace}
\newcommand{\PbPb}         {\mbox{Pb--Pb}\xspace}

\newcommand{\pt}           {\ensuremath{p_{\rm T}}\xspace}

\newcommand{\etarange}[1]  {\mbox{$\left | \eta \right |~<~#1$}}

\newcommand{\dEdx}         {\ensuremath{\textrm{d}E/\textrm{d}x}\xspace}

\newcommand{\nineH}        {$\sqrt{s}~=~0.9$~Te\kern-.1emV\xspace}
\newcommand{\seven}        {$\sqrt{s}~=~7$~Te\kern-.1emV\xspace}
\newcommand{\twoH}         {$\sqrt{s}~=~0.2$~Te\kern-.1emV\xspace}
\newcommand{\twosevensix}  {$\sqrt{s}~=~2.76$~Te\kern-.1emV\xspace}
\newcommand{\five}         {$\sqrt{s}~=~5.02$~Te\kern-.1emV\xspace}
\newcommand{\twosevensixnn}{$\sqrt{s_{\mathrm{NN}}}~=~2.76$~Te\kern-.1emV\xspace}
\newcommand{\fivenn}       {$\sqrt{s_{\mathrm{NN}}}~=~5.02$~Te\kern-.1emV\xspace}
\newcommand{\eightnn}       {$\sqrt{s_{\mathrm{NN}}}~=~8.16$~Te\kern-.1emV\xspace}

\newcommand{\GeVc}         {GeV/$c$\xspace}

\newcommand{\GeVmass}      {GeV/$c^2$\xspace}

\newcommand{\ITS}          {\rm{ITS}\xspace}

\newcommand{\ZDC}          {\rm{ZDC}\xspace}

\newcommand{\TPC}          {\rm{TPC}\xspace}

\newcommand{\VZEROA}       {\rm{V0A}\xspace}
\newcommand{\VZEROC}       {\rm{V0C}\xspace}

\newcommand{\pip}          {\ensuremath{\pi^{+}}\xspace}
\newcommand{\pim}          {\ensuremath{\pi^{-}}\xspace}

\newcommand{\kk}           {\ensuremath{\rm{K}^{+}\rm{K}^{-}}\xspace}

%% file: econfmacros.tex
\def\beq{\begin{equation}}
\def\eeq#1{\label{#1}\end{equation}}
\def\eeqn{\end{equation}}

\def\beqa{\begin{eqnarray}}
\def\eeqa#1{\label{#1}\end{eqnarray}}
\def\eeqan{\end{eqnarray}}

\let\bar=\overbar

\def\Dslash{\not{\hbox{\kern-4pt $D$}}}
\def\dslash{\not{\hbox{\kern-2pt $\del$}}}

\def\msb{{\bar{\ssstyle M \kern -1pt S}}}

